\documentclass[sn-mathphys,Numbered]{sn-jnl}% Math and Physical Sciences Reference Style
\usepackage{graphicx}%
\usepackage{multirow}%
\usepackage{amsmath,amssymb,amsfonts}%
\usepackage{amsthm}%
\usepackage{mathrsfs}%
\usepackage[title]{appendix}%
\usepackage{enumitem}
\usepackage{xcolor}%
\usepackage{textcomp}%
\usepackage{manyfoot}%
\usepackage{booktabs}%
\usepackage{algorithm}%
\usepackage{algorithmicx}%
\usepackage{algpseudocode}%
\usepackage{listings}%

\newcommand{\be}{\begin{equation}}
\newcommand{\bea}{\begin{eqnarray}}
\newcommand{\ee}{\end{equation}}
\newcommand{\eea}{\end{eqnarray}}
\def\1eq#1{Eq.~(\ref{#1})}

\def\2eqs#1#2{Eqs.~(\ref{#1}) and (\ref{#2})}
\def\3eqs#1#2#3{Eqs.~(\ref{#1}), (\ref{#2}) and (\ref{#3})}

\def\fig#1{Fig.~\ref{#1}}

\def\ie{{\it i.e.}, }
\def\eg{{\it e.g.}, }

\def\s#1{{\scriptscriptstyle #1}}

\newcommand{\Ls}{ \mathit{L}_{{sg}}}

\newcommand{\w}{{\cal W}}

\def\g{\Gamma}

\newcommand{\fatg}{{\rm{I}}\!\Gamma}

\newcommand{\Cfat}{{\mathbb C}}

\newcommand{\IW}{{\cal I}_{\w}}

\begin{document}

\title[Evidence of the Schwinger mechanism from lattice QCD]{Evidence of the Schwinger mechanism from lattice QCD}


\author*[1]{\fnm{Mauricio} \sur{Narciso Ferreira}}\email{ansonar@uv.es}

\affil*[1]{\orgdiv{Department of Theoretical Physics and IFIC}, \orgname{University of Valencia and CSIC}, \orgaddress{\postcode{E-46100},  \city{Valencia}, \country{Spain}}}


\abstract{

In quantum chromodynamics (QCD), gluons acquire a mass scale through the action of the Schwinger mechanism. This mass emerges as a result of the dynamical formation of massless bound-states of gluons which manifest as longitudinally coupled poles in the vertices. In this contribution, we show how the presence of these poles can be determined from lattice QCD results for the propagators and vertices. The crucial observation that allows this determination is that the Schwinger mechanism poles induce modifications, called ``displacements'', to the Ward identities (WIs) relating two- and three-point functions. Importantly, the displacement functions correspond precisely to the Bethe-Salpeter amplitudes of the massless bound-states. We apply this idea to the case of the three-gluon vertex in pure Yang-Mills SU(3). Using lattice results in the corresponding WI, we find an unequivocal displacement and show that it is consistent with the prediction based on the Bethe-Salpeter equation.

}




\maketitle

%--------------------------------------------------
\section{Introduction}\label{intro}

One of the most celebrated features of quantum chromodynamics (QCD)~\cite{Marciano:1977su} is the emergent hadron mass (EHM)~\cite{Roberts:2020udq, Roberts:2020hiw, Roberts:2021xnz, Roberts:2021nhw, Binosi:2022djx, Papavassiliou:2022wrb,Ding:2022ows,Roberts:2022rxm}, \ie the nonperturbative generation of massive hadrons out of fundamental fields, gluons and quarks, that are massless at the level of the Lagrangian. In this context, crucial signals of EHM have been revealed in the infrared behavior of the QCD propagators and vertices through the synergy between gauge-fixed lattice simulations \cite{Mandula:1987rh,Bowman:2002bm,Skullerud:2003qu,Cucchieri:2006tf,Ilgenfritz:2006he,Sternbeck:2006rd,Kamleh:2007ud,Cucchieri:2007md,Cucchieri:2007rg,Bogolubsky:2007ud,Cucchieri:2008qm,Cucchieri:2008fc,Cucchieri:2009zt,Cucchieri:2009kk,Bogolubsky:2009dc,Oliveira:2009eh,Cucchieri:2010mfr,Oliveira:2010xc,Boucaud:2011ug,Ayala:2012pb,Oliveira:2012eh,Sternbeck:2012mf,Bicudo:2015rma,Duarte:2016iko,Athenodorou:2016oyh,Duarte:2016ieu,Oliveira:2016muq,Boucaud:2017obn,Sternbeck:2017ntv,Boucaud:2018xup,Cucchieri:2018leo,Cucchieri:2018doy,Oliveira:2018lln,Dudal:2018cli,Vujinovic:2018nqc,Cui:2019dwv,Zafeiropoulos:2019flq,Aguilar:2019uob,Maas:2020zjp,Kizilersu:2021jen,Aguilar:2021lke,Aguilar:2021okw,Pinto-Gomez:2022brg,Pinto-Gomez:2022qjv,Pinto-Gomez:2023lbz} and continuum Schwinger function methods (CSM)~\cite{Qin:2020rad,Roberts:2020hiw,Cui:2020tdf,Chang:2021utv,Cui:2022bxn,Lu:2022cjx,Ding:2022ows,Roberts:2022rxm}, such
as Schwinger--Dyson equations (SDEs) \cite{Roberts:1994dr,Alkofer:2000wg,Fischer:2006ub,Roberts:2007ji,Binosi:2009qm,Bashir:2012fs,Binosi:2014aea,Cloet:2013jya,Aguilar:2015bud,Binosi:2016rxz,Binosi:2016nme,Huber:2018ned} and the functional renormalization group \cite{Pawlowski:2003hq, Pawlowski:2005xe,Fischer:2008uz,Carrington:2012ea,Carrington:2014lba,Cyrol:2017ewj, Corell:2018yil,Huber:2020keu,Dupuis:2020fhh, Blaizot:2021ikl,Pawlowski:2022oyq}. In particular, it is now established that the gluon propagator saturates to a finite value at the origin~\cite{Cucchieri:2007md,Cucchieri:2007rg,Bogolubsky:2007ud,Bogolubsky:2009dc,Oliveira:2009eh,Oliveira:2010xc,Cucchieri:2009zt,Cucchieri:2009kk,Cucchieri:2010mfr,Boucaud:2011ug,Sternbeck:2012mf,Oliveira:2012eh,Bicudo:2015rma,Kamleh:2007ud,Ayala:2012pb,Duarte:2016iko,Dudal:2018cli,Aguilar:2019uob,Horak:2022aqx}, which is an unequivocal signal of the dynamical generation of a gluon mass scale proposed decades ago~\cite{Cornwall:1979hz,Parisi:1980jy,Cornwall:1981zr,Bernard:1981pg,Bernard:1982my,Donoghue:1983fy}.

Gluon mass generation has far-reaching implications. For instance, it prevents QCD from developing a Landau pole, causes the effective decoupling of gluonic modes beyond a maximum gluon wavelength~\cite{Brodsky:2008be}, and suppresses Gribov copies~\cite{Braun:2007bx,Binosi:2014aea,Gao:2017uox}. Moreover, it sets a scale for many other dimensionful quantities, such as glueball masses~\cite{Meyers:2012ka,Sanchis-Alepuz:2015hma,Souza:2019ylx,Huber:2020ngt,Huber:2021yfy}. The importance of gluon mass generation has thus prompted an intense effort to elucidate the mechanism behind its dynamical origin.

The notion that gauge bosons can acquire masses dynamically, without violating gauge symmetry, originated with Schwinger in the sixties~\cite{Schwinger:1962tn,Schwinger:1962tp} and has been studied in various contexts since~\cite{Jackiw:1973tr,Jackiw:1973ha,Cornwall:1973ts,Eichten:1974et,Smit:1974je,Poggio:1974qs,Cornwall:1981zr,Papavassiliou:1989zd,Aguilar:2008xm,Aguilar:2011xe,Aguilar:2011ux,Ibanez:2012zk,Binosi:2012sj,Aguilar:2012rz,Aguilar:2016vin,Aguilar:2016ock,Eichmann:2021zuv,Aguilar:2021uwa,Aguilar:2022thg,Papavassiliou:2022wrb,Ferreira:2023fva}. In the particular case of QCD, the activation of the Schwinger mechanism for gluon mass generation hinges on the dynamical formation of massless, color-carrying, bound-states of gluons~\cite{Aguilar:2011xe,Ibanez:2012zk,Aguilar:2015bud,Aguilar:2016vin,Aguilar:2016ock,Aguilar:2021uwa,Aguilar:2022thg,Papavassiliou:2022wrb,Ferreira:2023fva}. Such massless bound-states appear as poles in the interaction vertices, which, in turn, lead to the saturation of the propagator.

A difficulty that arises in the quest to confirm the occurrence of the Schwinger mechanism in QCD is that lattice simulations can only compute transverse projections of the interaction vertices~\cite{Skullerud:2003qu,Cucchieri:2006tf,Cucchieri:2008qm,Athenodorou:2016oyh,Duarte:2016ieu,Boucaud:2017obn,Aguilar:2019uob,Aguilar:2021lke,Pinto-Gomez:2022brg}. However, the Schwinger mechanism poles in the vertices are strictly longitudinally coupled~\cite{Aguilar:2008xm,Ibanez:2012zk,Aguilar:2015bud,Aguilar:2016vin,Aguilar:2016ock,Aguilar:2021uwa,Aguilar:2022thg,Papavassiliou:2022wrb,Ferreira:2023fva}, and thus cannot be directly seen in lattice results for the vertex functions.

Recently~\cite{Aguilar:2021uwa,Aguilar:2022thg}, a method for confirming the existence of Schwinger mechanism poles from lattice QCD has been put forth. This method is based on the observation that the massless vertex poles induce crucial modifications to the Ward identities (WIs) relating propagators and vertices. These modifications, called ``displacements'', consist of the appearance of the Bethe-Salpeter (BS) amplitudes of the bound-states in the identities~\cite{Aguilar:2015bud,Aguilar:2016vin,Papavassiliou:2022wrb,Ferreira:2023fva}, in addition to the propagators and pole-free vertex parts that are present with or without the Schwinger mechanism. Hence, since the propagators and pole-free vertex parts are accessible to lattice simulations, the combination of lattice results for these quantities into the WIs allows us to determine the BS amplitude. Then, if the latter is found to be nonzero, the method allows us to confirm the occurrence of the Schwinger mechanism.

In the present contribution, we provide in Section~\ref{SM} a brief overview of the Schwinger mechanism and its realization in QCD through the formation of massless poles in the vertices. For simplicity, we neglect the effect of dynamical quarks, focusing instead on the pure Yang-Mills SU(3). Next, in Section~\ref{WIdisp} we illustrate through the case of an Abelian vertex how such massless poles displace the usual WIs. There we also present the WI displacement for the three-gluon vertex, which will allow us to determine the BS amplitude of the three-gluon vertex massless pole from lattice ingredients. Then, in Section~\ref{Wdet}, we discuss the determination of the function $\w(r)$, which is a special derivative of the ghost-gluon kernel and appears in the three-gluon vertex WI displacement. In Section~\ref{Cdet} we use the results of the previous sections to determine the BS amplitude, analyzing the statistical significance of the result and comparing it to the theoretical prediction obtained directly from the Bethe-Salpeter equation (BSE). Finally, in Section~\ref{conc} we present our conclusions.

%--------------------------------------------------
\section{Overview of the Schwinger mechanism}\label{SM}

In the Landau gauge, which will be used throughout this work, the gluon propagator can be written as $\Delta^{ab}_{\mu\nu}(q) = - i \delta^{ab}\Delta(q)P_{\mu\nu}(q)$, where $P_{\mu\nu}(q) := g_{\mu\nu} - q_\mu q_\nu/q^2$ is the transverse projector.

%%%%%%%%%%%%%%%%%%%%%%%%%%%%%%%%%%
%Fig. 1 - Gluon SDE
%%%%%%%%%%%%%%%%%%%%%%%%%%%%%%%%%%
\begin{figure}[H]
 \centering
 \includegraphics[width=1.0\textwidth]{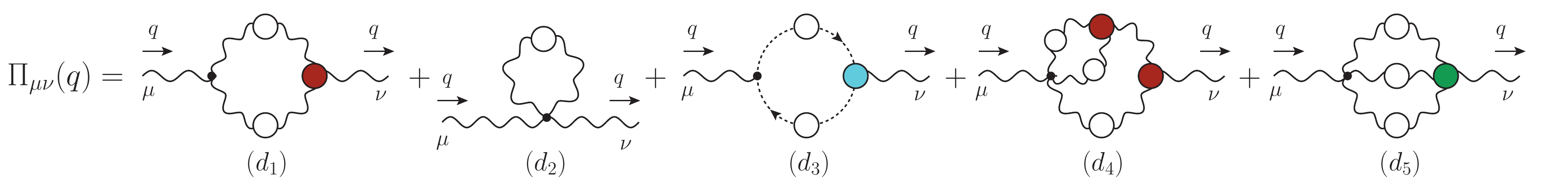}
\caption{Diagrammatic representation of the gluon self-energy, $\Pi_{\mu\nu}(q)$. In all our diagrams, wavy and dashed lines represent gluon and ghost fields, respectively, while circles denote dressed propagators and vertices. Feynman rules appropriate to our conventions are given in Appendix B of \cite{Binosi:2009qm}.}
\label{fig:SDEs}
\end{figure}
%%%%%%%%%%%%%%%%%%%%%%%%%%%%%%%%%%

\begin{figure}[H]
\centering
\raisebox{-0.5\height}{\includegraphics[width=0.55\textwidth]{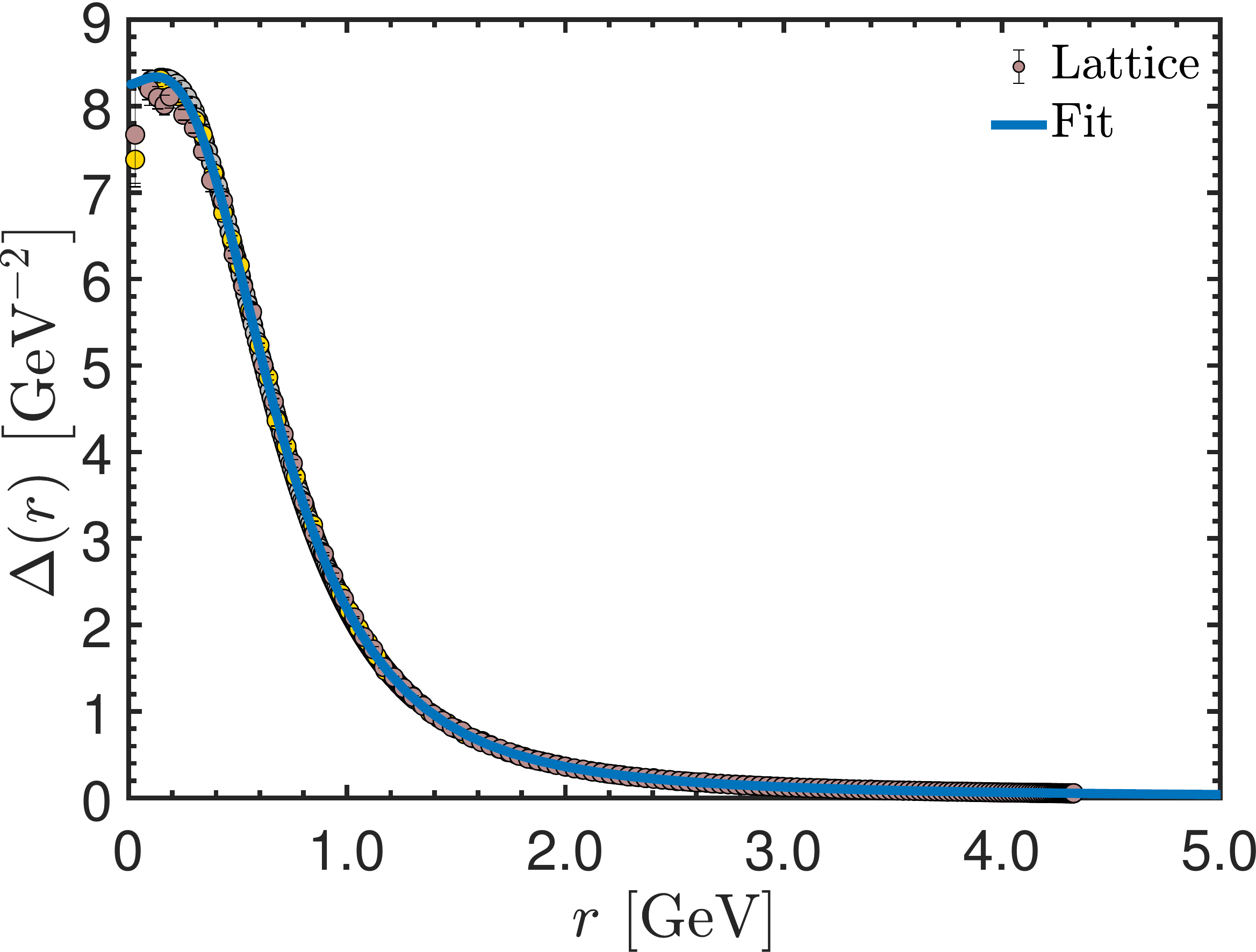}}
\caption{ Lattice data (points) from Ref.~\cite{Aguilar:2021okw} for the gluon propagator, compared to a physically motivated fit given by Eq.~(C11) of \cite{Aguilar:2021uwa} (blue solid).
}
\label{fig:gluon}
\end{figure}

The gluon propagator is determined in terms of the self-energy, $\Pi_{\mu\nu}(q)$, given diagrammatically in \fig{fig:SDEs}. Gauge symmetry requires that $\Pi_{\mu\nu}(q) = q^2 {\bf \Pi}(q) P_{\mu\nu}(q)$, where ${\bf \Pi}(q)$ defines the dimensionless vacuum polarization. Then,
\be 
\Delta^{-1}({q})=q^2 [1 + i {\bf \Pi}(q)]\,.
\label{vacpol}
\ee
The emergence of a gluon mass is signaled by the saturation of $\Delta(0)$ to a finite value, illustrated in \fig{fig:gluon} with recent lattice data from Ref.~\cite{Aguilar:2021okw}.

%%%%%%%%%%%%%%%%%%%%%%%%%%%%%%%%%%
% Figure 2 - lattice gluon propagator
%%%%%%%%%%%%%%%%%%%%%%%%%%%%%%%%%%
%%%%%%%%%%%%%%%%%%%%%%%%%%%%%%%%%%

The Schwinger mechanism is based on the observation that if the vacuum polarization acquires a pole at zero momentum transfer $\Delta(0)$ will saturate, even though no gluon mass term appears in the Lagrangian. Indeed, in the presence of such a pole \1eq{vacpol} has the limit
\be
\lim_{q \to 0} i{\bf \Pi}(q) = m^2/q^2 \,\,\Longrightarrow \,\,\lim_{q \to 0} \,\Delta^{-1}(q) = \lim_{q \to 0} \,(q^2 + m^2) \,\,\Longrightarrow \,\,\Delta^{-1}(0) = m^2\,,
\label{schmech}
\ee
written here in Euclidean space.

\begin{figure}[H]
 \centering
 \includegraphics[width=1\textwidth]{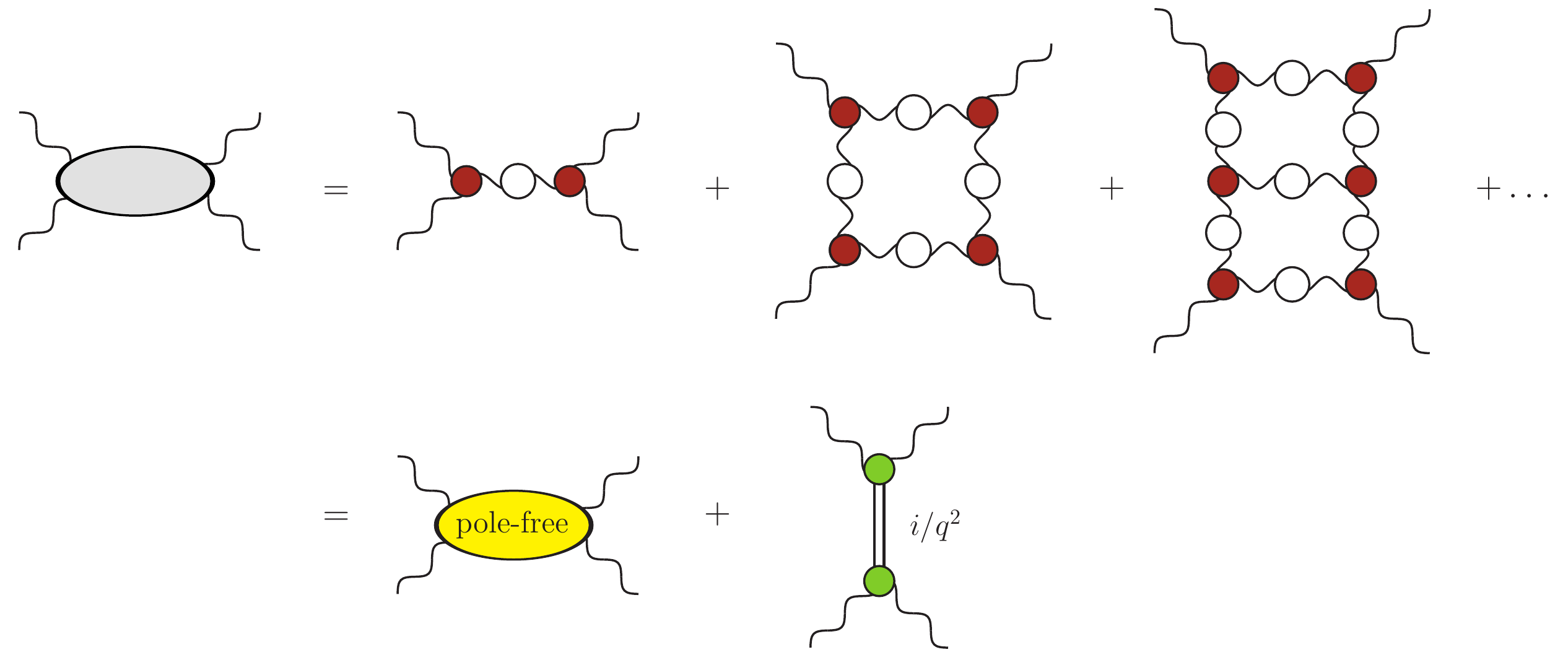}
\caption{Diagrammatic representation of the emergence of massless poles in the gluon-gluon scattering kernel. This kernel is one-particle irreducible with respect to vertical cuts. The first line shows some of the infinitely many diagrams contributing to the kernel. This tower of interactions is assumed to lead to the formation of a massless bound-state, with propagator $i/q^2$, such that the kernel acquires a pole-free and a pole contribution.}
\label{fig:4p_kernel}
\end{figure}

\begin{figure}[H]
 \centering
 \includegraphics[width=1\textwidth]{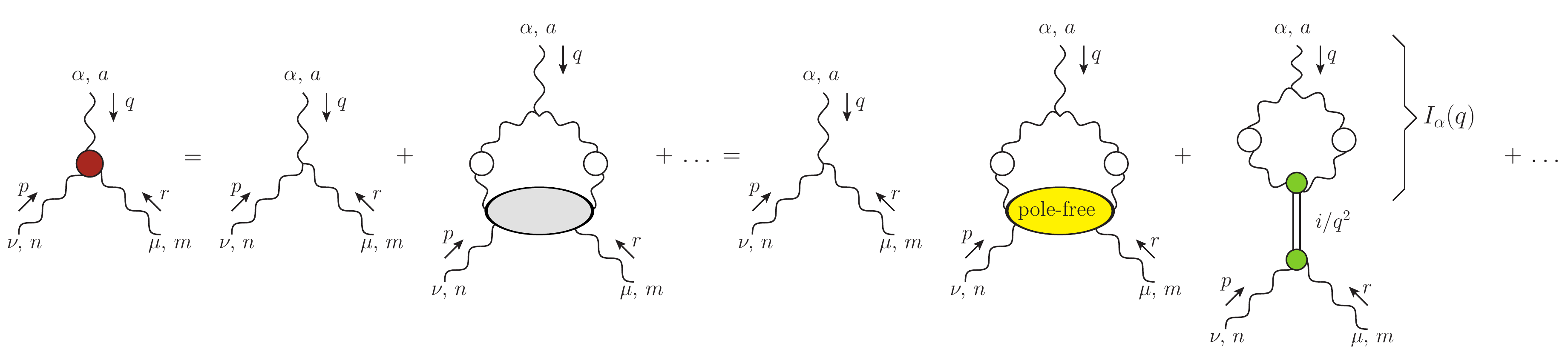}
\caption{Diagrammatic representation of the Schwinger-Dyson equation (SDE) for the three-gluon vertex, where appears the gluon-gluon scattering kernel of  \fig{fig:4p_kernel}. The bracket defines the amplitude, $I_\alpha(q)$, for a gluon to transition to a massless bound-state. Note that, by Bose symmetry, there must also exist poles in the $r$ and $p$ channels, which are not shown. The ellipsis denotes additional diagrams that are omitted for simplicity.}
\label{fig:3g_pole}
\end{figure}

The mechanism leading to the emergence of a pole in ${\bf \Pi}(0)$ can vary for different theories, see \eg \cite{Jackiw:1973tr,Jackiw:1973ha}. For Yang-Mills theories, an elegant nonperturbative mechanism has been put forward which is based on the formation of a special kind of bound-state of gluons~\cite{Eichten:1974et,Smit:1974je,Cornwall:1981zr,Papavassiliou:1989zd,Aguilar:2008xm,Aguilar:2011xe,Aguilar:2011yb,Aguilar:2011ux,Ibanez:2012zk,Binosi:2012sj,Aguilar:2015bud,Aguilar:2016vin,Aguilar:2016ock,Aguilar:2021uwa,Papavassiliou:2022wrb,Ferreira:2023fva}. This mechanism can be outlined through the following sequence of ideas: 
\begin{enumerate}[label=({\itshape\roman*})]
\item First, it is assumed that the gluon self-interaction is strong enough to form massless colored bound-states. These bound-states can be shown to not appear in $S$-matrix elements, such that no new massless particle is introduced in the spectrum of the theory~\cite{Jackiw:1973tr,Jackiw:1973ha,Eichten:1974et,Smit:1974je,Aguilar:2011xe,Papavassiliou:2022wrb}. Nevertheless, the bound-state propagator, $i/q^2$, induces a pole in a certain gluon-gluon scattering kernel, illustrated diagrammatically in \fig{fig:4p_kernel}.

\item Consequently, the fundamental vertices of the theory acquire poles at zero momentum transfer. This can be clearly seen in the case of the three-gluon vertex by analyzing the SDE that governs its momentum evolution, shown in \fig{fig:3g_pole}. Indeed, in that equation appears the aforementioned gluon-gluon scattering kernel, which induces a pole in the vertex.

\item Finally, the massless poles in the vertices make their way naturally into the vacuum polarization, thus activating the Schwinger mechanism.
\end{enumerate}
%

%%%%%%%%%%%%%%%%%%%%%%%%%%%%%%%%%%
%Fig. 3 - Gluon-gluon scattering kernel
%%%%%%%%%%%%%%%%%%%%%%%%%%%%%%%%%%
%%%%%%%%%%%%%%%%%%%%%%%%%%%%%%%%%%

%%%%%%%%%%%%%%%%%%%%%%%%%%%%%%%%%%
%Fig. 4 - 3g pole
%%%%%%%%%%%%%%%%%%%%%%%%%%%%%%%%%%
%%%%%%%%%%%%%%%%%%%%%%%%%%%%%%%%%%

From now on we will focus on the three-gluon vertex, whose associated massless bound-state pole is expected to be the leading contributor to gluon mass generation~\cite{Aguilar:2008xm,Aguilar:2013hoa,Aguilar:2017dco,Aguilar:2021uwa,Aguilar:2022thg}. We denote this vertex by $\fatg^{amn}_{\alpha\mu\nu}(q,r,p) = g f^{amn}\fatg_{\alpha\mu\nu}(q,r,p)$, where $g$ is the gauge coupling and $f^{amn}$ are the SU(3) structure constants. 

The emergence of massless bound-state transitions in the three-gluon vertex prompts us to split $\fatg_{\alpha\mu\nu}(q,r,p)$ into a pole-free part, $\g_{\alpha\mu\nu}(q,r,p)$, and a pole contribution, $V_{\alpha\mu\nu}(q,r,p)$, \ie
\be
\fatg_{\alpha\mu\nu}(q,r,p) = \g_{\alpha\mu\nu}(q,r,p) + V_{\alpha\mu\nu}(q,r,p) \,. \label{split3g}
\ee

The dynamical origin of $V_{\alpha\mu\nu}(q,r,p)$ in the formation of massless poles imposes a crucial constraint on its Lorentz structures. Specifically, by Lorentz symmetry, the amplitude $I_{\alpha}(q)$ for a gluon to transition to a massless bound-state, defined by the bracket in \fig{fig:3g_pole}, must be of the form $I_\alpha(q) = q_\alpha I(q)$, for some scalar $I(q)$. Hence, the $q = 0$ pole in the vertex must be associated with tensor structures longitudinal to the leg carrying momentum $q$. Similar considerations show that the poles at $r = 0$ and $p = 0$ must be associated with tensors longitudinal to $r_\mu$ and $p_\nu$, respectively. Therefore, the massless poles that trigger the Schwinger mechanism must be strictly longitudinally coupled, such that~\cite{Aguilar:2011xe,Ibanez:2012zk}
\be
\label{eq:transvp}
{P}_{\alpha'}^{\alpha}(q){P}_{\mu'}^{\mu}(r){P}_{\nu'}^{\nu}(p) V_{\alpha\mu\nu}(q,r,p) = 0 \,.
\ee

From \1eq{eq:transvp}, together with Bose symmetry of the vertex, we see that the pole part $V_{\alpha\mu\nu}(q,r,p)$ can be written as
\be 
V_{\alpha\mu\nu}(q,r,p) = \left(\frac{q_\alpha}{q^2}\right)C_{\mu\nu}(q,r,p) + \left(\frac{r_\mu}{r^2}\right)C_{\nu\alpha}(r,p,q) + \left(\frac{p_\nu}{p^2}\right)C_{\alpha\mu}(p,q,r) \,, \label{V_Lorentz}
\ee
with
\be 
C_{\mu\nu}(q,r,p) = C_1 g_{\mu\nu} + C_2 r_\mu r_\nu + C_3 p_\mu p_\nu + C_4 r_\mu p_\nu + C_5 p_\mu r_\nu \,, \label{theCs}
\ee
where $C_i \equiv C_i(q,r,p)$. Due to the transversality of the Landau gauge gluon propagator, the form factors $C_{2,3,4}$ decouple in most calculations. Moreover, in this gauge, the form factor $C_5$ can be shown to not contribute to the gluon mass~\cite{Aguilar:2011xe,Ibanez:2012zk,Aguilar:2021uwa}. Hence, we will restrict our discussion to $C_1$.

\begin{figure}[H]
\centering
\raisebox{-0.5\height}{\includegraphics[width=0.55\textwidth]{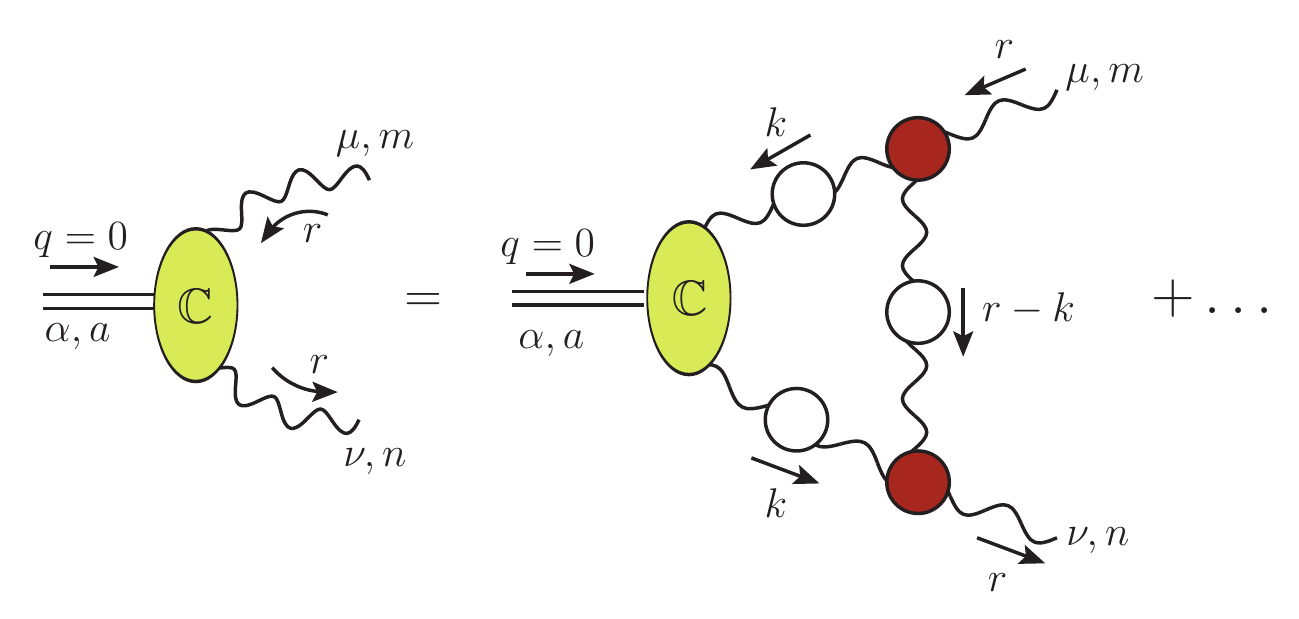}}\hfil\raisebox{-0.5\height}{\includegraphics[width=0.44\textwidth]{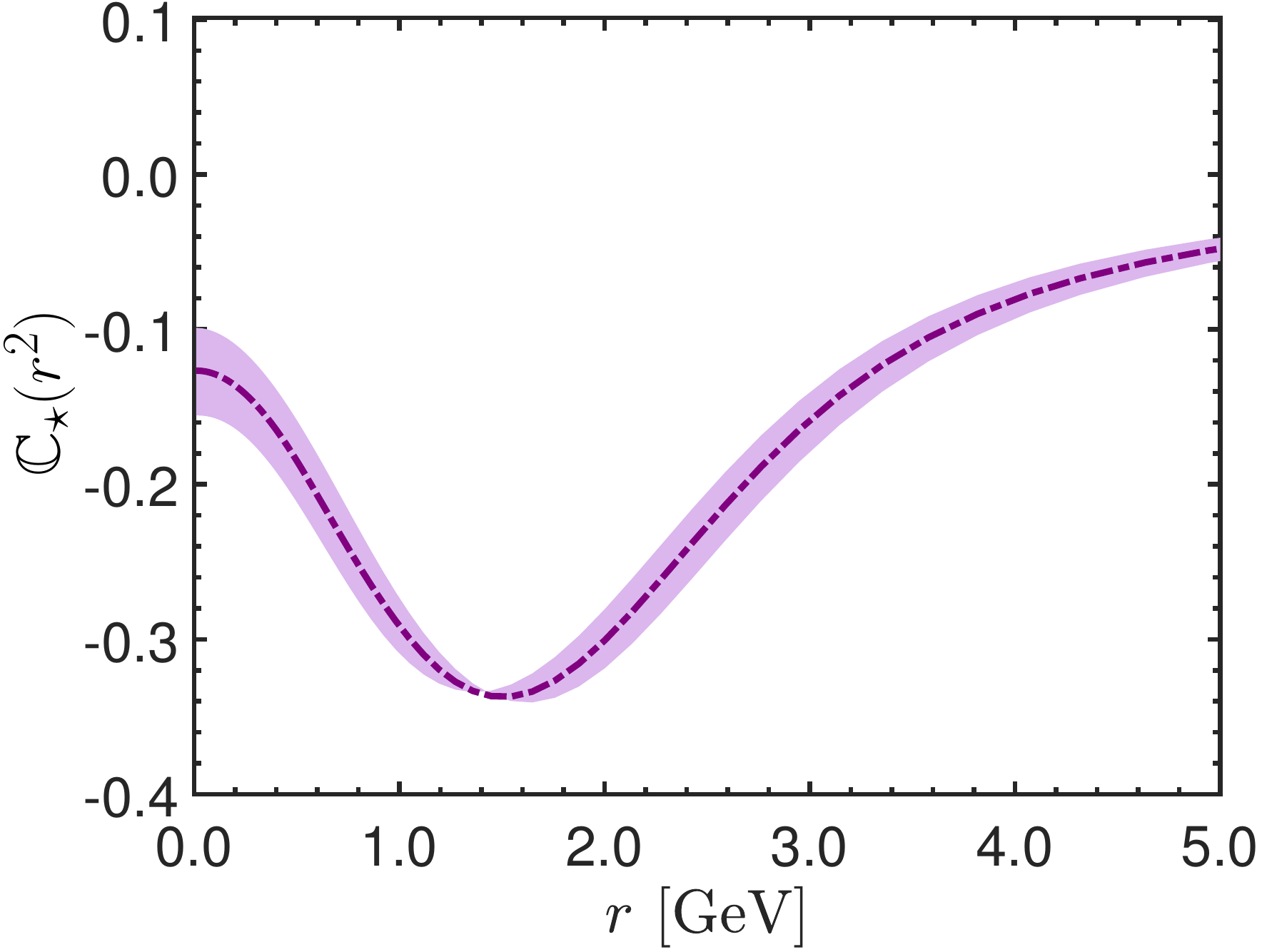}}
\caption{ Left: Bethe-Salpeter equation (BSE) governing the formation of the massless bound-state that triggers the Schwinger mechanism. The ellipsis denotes higher-order corrections to the gluon-gluon scattering kernel and coupling to poles in vertices other than the three-gluon~\cite{Aguilar:2011xe,Ibanez:2012zk,Aguilar:2017dco,Aguilar:2021uwa,Papavassiliou:2022wrb,Ferreira:2023fva}. Right: Bethe-Salpter (BS) amplitude, $\Cfat(r)$, obtained in \cite{Aguilar:2021uwa}, using the BSE of the left panel.
}
\label{fig:BSE}
\end{figure}

At this point, we emphasize that the massless bound-state that triggers the Schwinger mechanism in QCD is not put in by hand, but emerges dynamically. Indeed, as with any other bound-state, its formation is governed by a BSE~\cite{Aguilar:2011xe,Ibanez:2012zk,Aguilar:2017dco,Aguilar:2021uwa,Papavassiliou:2022wrb,Ferreira:2023fva}, represented diagrammatically in the left panel of \fig{fig:BSE}.

The function that plays the role of BS amplitude in the BSE of \fig{fig:BSE} is denoted by $\Cfat(r)$ and is related to the form factor $C_1$  defined in \2eqs{V_Lorentz}{theCs}. Specifically, note that Bose symmetry requires $C_1(q,r,p) = - C_1(q,p,r)$, such that 
\be 
C_1(0,r,-r) = 0\,. \label{C0}
\ee
Then, in the vicinity of the $q = 0$ pole,
\be 
C_1(q,r,p) = 2(q\cdot r)\Cfat(r) \,, \qquad \Cfat(r) := \left.\frac{\partial C_1(q,r,p)}{\partial p^2}\right\vert_{q = 0} \,. \label{Cfat_def}
\ee
%

%%%%%%%%%%%%%%%%%%%%%%%%%%%%%%%%%%
% Figure 5 - BSE
%%%%%%%%%%%%%%%%%%%%%%%%%%%%%%%%%%
%%%%%%%%%%%%%%%%%%%%%%%%%%%%%%%%%%

The vital first test of the Schwinger mechanism is the existence of nontrivial solutions for $\Cfat(r)$. Indeed, previous studies have shown that the BSE of \fig{fig:BSE} admits nontrivial solutions, using lattice inputs for the propagator and three-gluon vertex therein~\cite{Aguilar:2011xe,Ibanez:2012zk,Aguilar:2017dco,Aguilar:2021uwa,Papavassiliou:2022wrb,Ferreira:2023fva}. The most up-to-date solution was obtained in Ref.~\cite{Aguilar:2021uwa} and is shown in the right panel of \fig{fig:BSE}. For later convenience, this solution is denoted by $\Cfat_\star(r)$, to distinguish it from the $\Cfat(r)$ that will be determined in Section~\ref{Cdet} from the WI displacement. Note that, since the BSE of \fig{fig:BSE} is a homogeneous equation, it only determines $\Cfat_\star(r)$ up to a multiplicative constant; the particular solution shown there has its scale set by matching it to the result obtained in Section~\ref{Cdet}, as explained therein.

To conclude this section, we remark that coupling the poles of the three-gluon and ghost-gluon vertices does not significantly affect the solution shown \fig{fig:BSE}. Moreover, the pole associated with the ghost-gluon vertex is subleading in comparison to $\Cfat(r)$~\cite{Aguilar:2017dco,Aguilar:2021uwa}.

\section{Ward identity displacement}\label{WIdisp}

It follows from the longitudinality property of $V_{\alpha\mu\nu}(q,r,p)$, \ie from \1eq{eq:transvp}, that the Schwinger mechanism massless poles cannot be computed on the lattice by direct simulation of the three-gluon vertex. Indeed, lattice QCD can only determine the transverse projections of the vertex functions. In particular, for the three-gluon vertex, lattice observables involve the projection ${\overline \g}_{\alpha\mu\nu}(q,r,p)$, defined by~\cite{Cucchieri:2006tf,Cucchieri:2008qm,Athenodorou:2016oyh,Duarte:2016ieu,Boucaud:2017obn,Aguilar:2019uob,Aguilar:2021lke,Pinto-Gomez:2022brg}
\be 
{\overline \g}_{\alpha\mu\nu}(q,r,p) := P^{\alpha'}_\alpha(q)P^{\mu'}_\mu(r)P^{\nu'}_\nu(p)\fatg_{\alpha'\mu'\nu'}(q,r,p) \,, \label{3gtransv}
\ee
rather than $\fatg_{\alpha'\mu'\nu'}(q,r,p)$ itself.
Then, using \2eqs{split3g}{eq:transvp}, we see that
\be 
{\overline \g}_{\alpha\mu\nu}(q,r,p) = P^{\alpha'}_\alpha(q)P^{\mu'}_\mu(r)P^{\nu'}_\nu(p)\g_{\alpha'\mu'\nu'}(q,r,p) \,,
\ee
\ie lattice simulations only have access to the pole-free part of the vertex.

Nevertheless, a method for determining the BS amplitude, $\Cfat(r)$, from lattice results has recently been devised~\cite{Aguilar:2021uwa,Aguilar:2022thg}. The crucial observation that enables this determination is that the BS amplitudes of the massless vertex poles appear in the WI which relate two and three-point sector functions~\cite{Aguilar:2015bud,Aguilar:2016vin,Papavassiliou:2022wrb,Ferreira:2023fva}.

To fix the ideas, consider for simplicity the ghost-gluon vertex in the background field method~\cite{DeWitt:1967ub,tHooft:1971qjg,Honerkamp:1972fd,Kallosh:1974yh,Kluberg-Stern:1974nmx,Arefeva:1974jv,Abbott:1980hw,Weinberg:1980wa,Abbott:1981ke,Shore:1981mj,Abbott:1983zw}, denoted by  ${\widetilde \g}_{\mu}(q,r,p)$, where $q$, $r$ and $p$ stand for the gluon, antighost, and ghost momenta, respectively. This vertex satisfies a Slavnov--Taylor identity (STI)~\cite{Taylor:1971ff,Slavnov:1972fg} identical in form to that of the photon-scalar vertex of scalar QED. Specifically~\cite{Cornwall:1989gv,Aguilar:2006gr,Binosi:2009qm},
\be 
q^\mu {\widetilde \g}_{\mu}(q,r,p) = D^{-1}(p) - D^{-1}(r) \,, \label{abelian_STI}
\ee
where $D^{ab}(q) = i\delta^{ab}D(q)$ denotes the ghost propagator. Note that, at tree level ${\widetilde \g}_{\mu}(q,r,p) = ( r - p )_\mu$.

Now, let us assume that ${\widetilde \g}_{\mu}(q,r,p)$ is a pole-free function at $q = 0$. From \1eq{abelian_STI}, we can derive the textbook WI by expanding both sides to the first order in $q = 0$ and equating coefficients of equal orders. This procedure yields,
\be 
{\widetilde \g}_{\mu}(0,r,-r)
= 2 r_\mu \frac{ \partial D^{-1}(r) }{\partial r^2} \,. \label{Abelian_WI_tens}
\ee
Equivalently, since Lorentz invariance implies ${\widetilde \g}_{\mu}(0,r,-r) = r_\mu {\widetilde {\cal A}}(r)$, for some scalar function ${\widetilde {\cal A}}(r)$, \1eq{Abelian_WI_tens} can be recast as
\be 
{\widetilde {\cal A}}(r) = 2\frac{ \partial D^{-1}(r) }{\partial r^2} \,. \label{Abelian_WI}
\ee

Next, let us activate the Schwinger mechanism, such that the vertex acquires a pole at $q = 0$. By analogy to \1eq{split3g}, we write
\be 
{\widetilde \g}_{\mu}(q,r,p) \to {\widetilde \fatg}_{\mu}(q,r,p) = {\widetilde \g}_{\mu}(q,r,p) + \frac{q_\mu}{q^2}{\widetilde C}(q,r,p) \,, \label{Abelian_split}
\ee
where ${\widetilde \g}_{\mu}(q,r,p)$ now represents the pole-free part of the vertex only, while ${\widetilde C}(q,r,p)$ is the residue of the Schwinger mechanism pole.

Since the gauge symmetry is assumed to be unbroken, the STI of \1eq{abelian_STI} remains valid for the full vertex, \ie
\be 
q^\mu {\widetilde \fatg}_{\mu}(q,r,p) = q^\mu {\widetilde \g}_{\mu}(q,r,p) + {\widetilde C}(q,r,p) = D^{-1}(p) - D^{-1}(r) \,.\label{abelian_STI_poles}
\ee
Then we repeat the procedure of the derivation of the WI, expanding \1eq{abelian_STI_poles} in a Taylor series around $q = 0$. At zeroth order, \1eq{abelian_STI_poles} implies
\be 
{\widetilde C}(0,r,-r) = 0 \,,
\ee
which is akin to the \1eq{C0}, derived for the three-gluon vertex from Bose symmetry in Section~\ref{SM}.

Next, at first order we obtain
\be 
{\widetilde {\cal A}}(r) = 2\left[ \frac{ \partial D^{-1}(r) }{\partial r^2} - {\widetilde {\cal C} }(r) \right]\,, \qquad {\widetilde {\cal C} }(r) := \left. \frac{\partial {\widetilde C}(q,r,p)}{\partial p^2}\right\vert_{q = 0} \,. \label{Abelian_WI_poles}
\ee
Comparing \2eqs{Abelian_WI}{Abelian_WI_poles}, we see that the WI for the form factor ${\widetilde {\cal A}}(r)$ gets modified, or ``displaced'', by a derivative, ${\widetilde{\cal C}}(r)$, of the pole residue ${\widetilde C}(q,r,p)$. Note that the above definition for ${\widetilde{\cal C}}(r)$ is completely analogous to the BS amplitude $\Cfat(r)$ of the three-gluon vertex, defined in \1eq{Cfat_def}.

With \1eq{Abelian_WI_poles} at hand, if the propagator $D(r)$ and the vertex form factor ${\widetilde {\cal A}}(r)$ are somehow known, we can compute ${\widetilde{\cal C}}(r)$, thus determining if the vertex has a massless bound-state pole.

The same idea can be applied to the three-gluon vertex. The only fundamental difference is the non-Abelian nature of its STI, which implies that the relevant WI and its displacement have a more complicated form that mixes gluon and ghost sector functions.

\begin{figure}[H]
\centering
\includegraphics[width=0.475\textwidth]{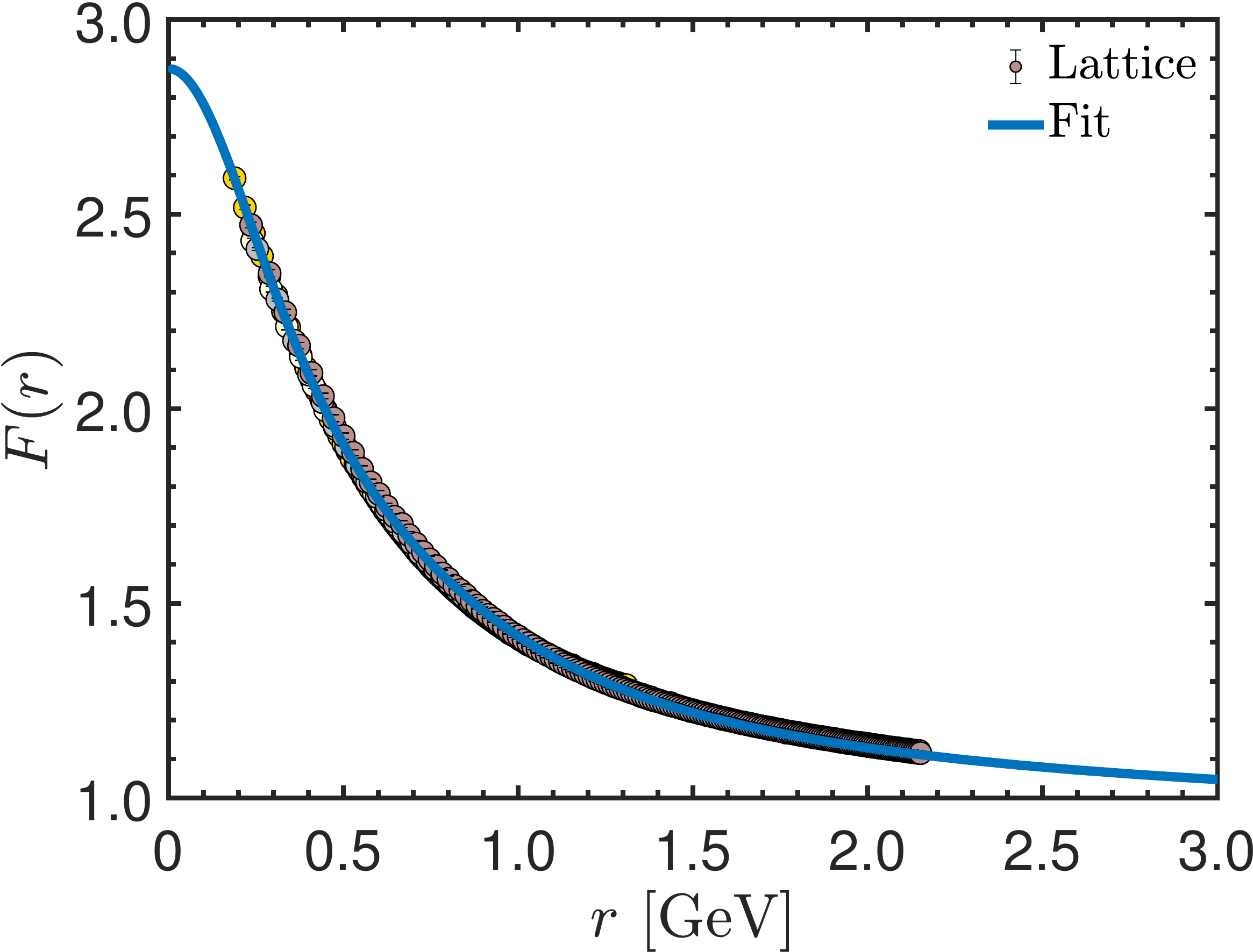}\hfil\includegraphics[width=0.475\textwidth]{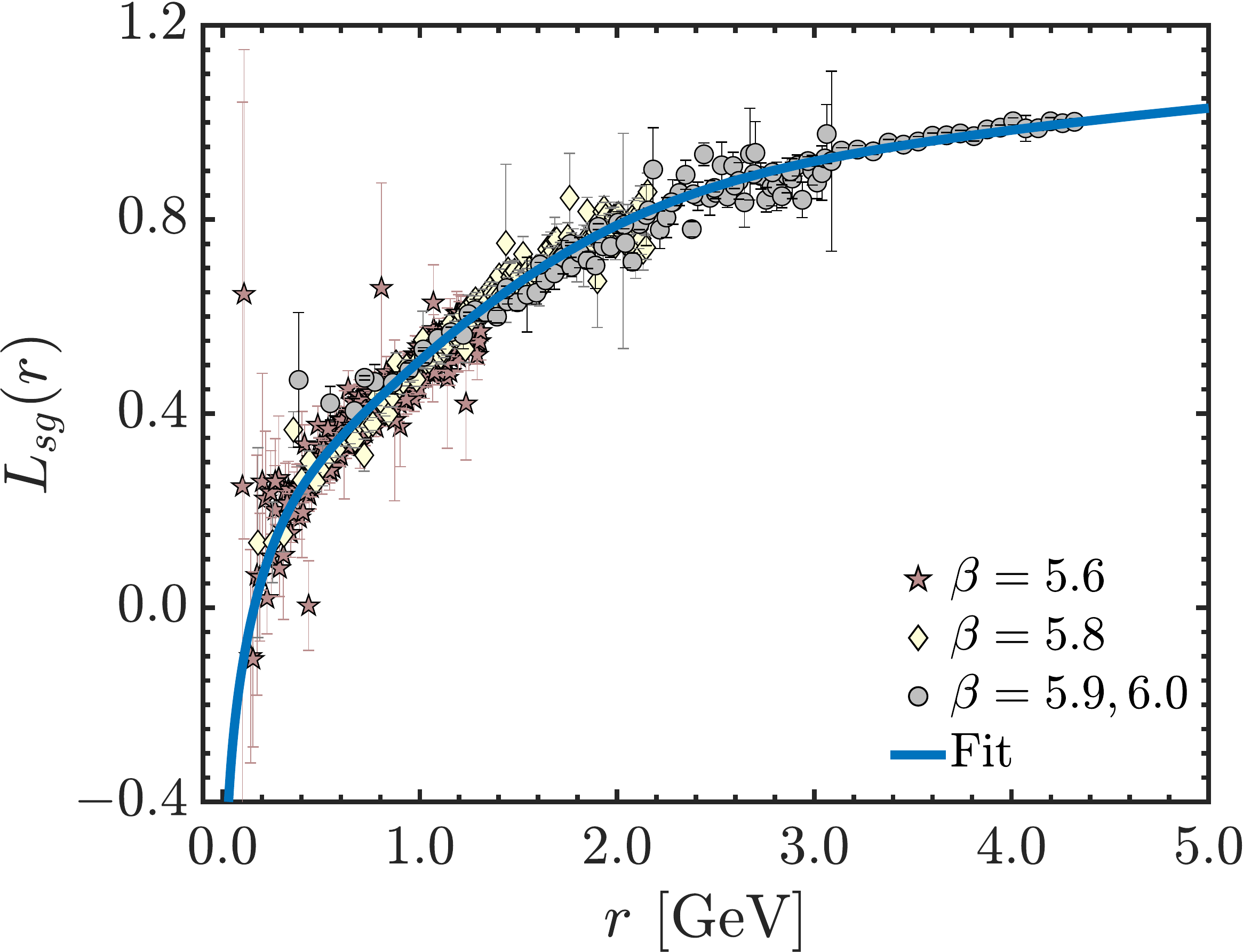}
\caption{ Left: Lattice data (points) from Refs.~\cite{Boucaud:2018xup,Aguilar:2021okw} for the ghost dressing function, $F(r)$. Right: Form factor $\Ls(r)$ of the three-gluon vertex in the soft gluon limit obtained from lattice quantum chromodynamics (QCD) in Ref.~\cite{Aguilar:2021lke} (points). The blue solid curve in each panel denotes a fit to the corresponding data, given by Eqs.~(C6) and (C12) of \cite{Aguilar:2021uwa}, for $F(r)$ and $\Ls(r)$, respectively.
}
\label{fig:Lsg}
\end{figure}

Specifically, the STI which relates the three-gluon vertex to the gluon propagator is given by~\cite{Marciano:1977su,Ball:1980ax,Davydychev:1996pb,vonSmekal:1997ern,Binosi:2011wi,Gracey:2019mix}
\be
q^\alpha \fatg_{\alpha \mu \nu}(q,r,p) = F(q)
\left[\Delta^{-1}(p) P_\nu^\sigma(p) H_{\sigma\mu}(p,q,r) - \Delta^{-1}(r) P_\mu^\sigma(r) H_{\sigma\nu}(r,q,p)\right]\,,
\label{st1_conv} 
\ee
where $F(q)$ is the ghost dressing function, defined by $D(q) = F(q)/q^2$, and $H_{\nu\mu}(r,p,q)$ is the ghost-gluon kernel~\cite{Aguilar:2018csq}, which will be discussed in the next section. We point out that the ghost propagator remains massless, while its dressing function, $F(q)$, becomes finite at the origin~\cite{Aguilar:2008xm,Dudal:2008sp,Boucaud:2008ky,Boucaud:2008ji,Kondo:2009gc,Boucaud:2011ug,Pennington:2011xs,Dudal:2012zx,Aguilar:2013xqa,Cyrol:2016tym,Huber:2018ned,Aguilar:2018csq,Aguilar:2021okw,Ilgenfritz:2006he,Cucchieri:2007md,Bogolubsky:2007ud,Cucchieri:2008fc,Bogolubsky:2009dc,Ayala:2012pb,Boucaud:2018xup,Cui:2019dwv}, as shown in the left panel of \fig{fig:Lsg}.

%%%%%%%%%%%%%%%%%%%%%%%%%%%%%%%%%%
% Figure 6 - F(r) and Lsg(r)
%%%%%%%%%%%%%%%%%%%%%%%%%%%%%%%%%%
%%%%%%%%%%%%%%%%%%%%%%%%%%%%%%%%%%

The WI for the three-gluon vertex is obtained as a special case of the above STI. To derive it, one expands \1eq{st1_conv} around $q = 0$ and matches coefficients of equal orders on each side of the resulting equation. Evidently, the zeroth-order expansion leads again to \1eq{C0}. As for the first-order term, after a suitable projection to isolate the classical tensor structure of the three-gluon vertex, one obtains the relation (for detailed derivations see~\cite{Aguilar:2021uwa,Papavassiliou:2022wrb})
\be
\Cfat(r) = \Ls(r) - F(0)\left\{\frac{\w(r)}{r^2}\Delta^{-1}(r) + \widetilde{Z}_1 \frac{d\Delta^{-1}(r)}{dr^2} \right\} \,.
\label{centeuc}
\ee

In the above equation, the displacement of the WI is \emph{precisely} the BS amplitude $\Cfat(r)$ of the Schwinger pole of the three-gluon vertex. On the other hand, $\Ls(r)$ is the classical form factor of the three-gluon vertex in the soft gluon limit, defined by \cite{Aguilar:2021lke}
\be 
\Ls(r) = \left. \frac{\g_0^{\alpha\mu\nu}(q,r,p)P_{\alpha\alpha'}(q)P_{\mu\mu'}(r)P_{\nu\nu'}(p)\fatg^{\alpha'\mu'\nu'}(q,r,p)}{\g_0^{\alpha\mu\nu}(q,r,p)P_{\alpha\alpha'}(q)P_{\mu\mu'}(r)P_{\nu\nu'}(p)\g_0^{\alpha'\mu'\nu'}(q,r,p)} \right\vert_{q\to 0} \,,\label{Lsg_def}
\ee 
with $\g_0^{\alpha\mu\nu}(q,r,p)$ denoting the tree-level form of the vertex. By now, this form factor has been extensively studied on the lattice~\cite{Athenodorou:2016oyh,Duarte:2016ieu,Boucaud:2017obn,Aguilar:2019uob,Aguilar:2021lke,Pinto-Gomez:2022brg}, such that its form is rather accurately known. In the right panel of \fig{fig:Lsg} we show the lattice results for $\Ls(r)$ from \cite{Aguilar:2021lke} (points), together with a physically motivated fit for it given by Eq.~(C12) of \cite{Aguilar:2021uwa} (blue continuous curve).

Lastly, $\w(r)$ is a particular derivative of the ghost-gluon kernel, namely \cite{Aguilar:2020yni,Aguilar:2021uwa}
\be 
\w(r) = - \frac{1}{3r^2}P^{\mu\nu}(r)\left[\frac{\partial H_{\nu\mu}(p,q,r)}{\partial q^\alpha } \right]_{q=0} \,, \label{HKtens}
\ee 
while ${\widetilde Z}_1$ is the renormalization constant of the ghost-gluon vertex. The latter is finite in the Landau gauge, by virtue of the well-known Taylor theorem~\cite{Taylor:1971ff}.

The \1eq{centeuc} is the central relation that will enable us to determine the BS amplitude, $\Cfat(r)$, from lattice data for the gluon and ghost propagators and the form factor $\Ls(r)$ of the three-gluon vertex. To this end, we need first to determine the ghost-gluon kernel derivative $\w(r)$ appearing in \1eq{centeuc}.

%--------------------------------------------------
\section{Ghost-gluon kernel contribution}\label{Wdet}

\begin{figure}[H]
 \centering
 \includegraphics[width=\textwidth]{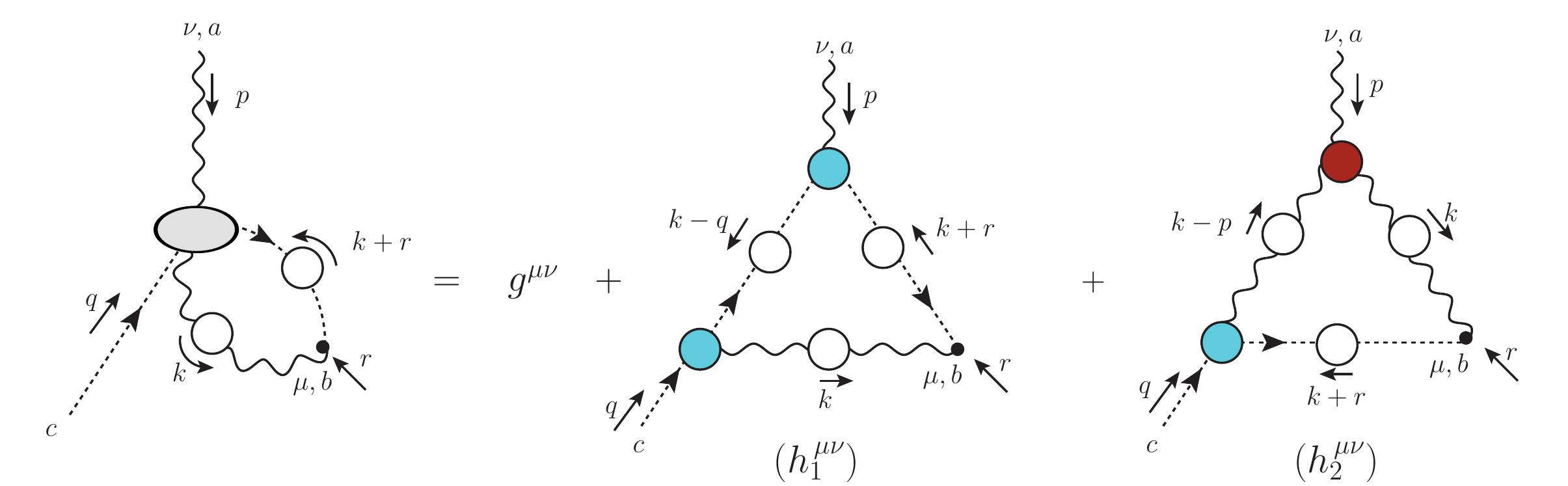}	
 \caption{ SDE for the ghost--gluon scattering kernel, $H_{\mu\nu}(r,q,p)$. We omit a diagram containing a 1PI four-point function, which has been shown to contribute to the ghost-gluon vertex at the $2\%$ level only~\cite{Huber:2017txg}. }
\label{fig:H_SDE}
\end{figure}

Now we briefly describe our lattice-driven SDE determination of $\w(r)$. The starting point of this analysis is the SDE that defines the ghost-gluon kernel, shown diagrammatically in \fig{fig:H_SDE}.

%%%%%%%%%%%%%%%%%%%%%%%%%%%%%%%%%%
% Figure 7 - H_munu SDE
%%%%%%%%%%%%%%%%%%%%%%%%%%%%%%%%%%
%%%%%%%%%%%%%%%%%%%%%%%%%%%%%%%%%%

From that equation, the function $\w(r)$ can be isolated through \1eq{HKtens}. The resulting expression for $\w(r)$ can be written as
\be
\w(r) = \w_1(r) + \w_2(r) \,, \label{W_conts}
\ee 
where the $\w_i(r)$ denote the contributions of the diagrams $(h_i^{\mu\nu})$ in \fig{fig:H_SDE}, respectively. These are given by
\begin{eqnarray}
\w_1(r) &=& {\widetilde \lambda} \int_k \Delta(k) D(k) D(k+r) (r\cdot k ) f(k,r)B_1( k+r, - k , -r )B_1(k,0,-k) \,, \nonumber\\
\w_2(r) &=& {\widetilde \lambda} \int_k \Delta(k) \Delta(k+r) D(k+r) B_1(k+r,0,-k-r) \IW(-r, -k, k+r) \,, \label{W_diags}
\end{eqnarray}
where ${\widetilde \lambda} := i g^2 C_{\rm A} {\widetilde Z}_1/6$, $C_\mathrm{A}$ is the Casimir eigenvalue of the adjoint representation [$N$ for SU$(N)$], and
\be 
f(k,r) := 1 - \frac{ (r \cdot k)^2 }{ r^2 k^2 } \,. \label{fqk_def}
\ee 

In addition to the gluon and ghost propagators, $\Delta(r)$ and $D(r)$, respectively, \1eq{W_diags} involves quantities that are related to the ghost-gluon and three-gluon vertices, namely $B_1(r,p,q)$, ${\widetilde Z}_1$ and ${\IW}(q,r,p)$. Below we explain their meaning in detail.
\begin{enumerate}[label=({\itshape\roman*})]
\item In \1eq{W_diags}, $B_1(r,p,q)$ denotes the classical form factor of the ghost-gluon vertex, $\fatg_\mu(r,p,q)$, whose most general tensor structure is given by
\be 
\fatg_\mu(r,p,q) = B_1(r,p,q) r_\mu + B_2(r,p,q) q_\mu \,.
\ee
Hence, at tree level $B_1^0 = 1$ and $B_2^0 = 0$. 

Note that the ghost-gluon vertex and kernel are related by the STI
\be 
\fatg_\mu(r,p,q) = r^\nu H_{\nu\mu}(q,r,p) \,. \label{H_to_Gamma}
\ee
Thus, the general kinematics $B_1(r,p,q)$ can be determined through another projection of the SDE of \fig{fig:H_SDE}. 

Such an SDE determination of the general kinematics $B_1(r,p,q)$ was performed in Refs.~\cite{Aguilar:2022thg,Ferreira:2023fva}, also using lattice results as inputs for all its ingredients. It is beyond the scope of the present work to describe this analysis in detail. It suffices to mention that the results for $B_1(r,p,q)$ deviate only moderately from its tree-level value, in agreement with several previous continuum studies~\cite{Schleifenbaum:2004id,Huber:2012kd,Aguilar:2013xqa,Cyrol:2016tym,Mintz:2017qri,Aguilar:2018csq,Huber:2018ned,Aguilar:2019jsj,Huber:2020keu,Barrios:2020ubx}, and reproduce the available lattice data from Ref.~\cite{Ilgenfritz:2006he,Sternbeck:2006rd}. As such, the impact of the precise dressing of $B_1(r,p,q)$ on the $\w(r)$ computed through \1eq{W_conts} is under stringent control.

\item As previously mentioned, the ghost-gluon kernel, and hence $\w(r)$, is finite in Landau gauge~\cite{Taylor:1971ff}. Nevertheless, multiplicative renormalization of the theory leads to the appearance~\cite{Aguilar:2020yni} of the ghost-gluon renormalization constant, ${\widetilde Z}_1$, in \1eq{W_diags}. The \emph{finite} value of this constant depends on the renormalization scheme adopted. 

To take the most advantage of the lattice data for the propagators and the three-gluon vertex, we adopt the scheme where $\Delta(r)$, $F(r)$ and $\Ls(r)$ are most readily renormalized. Namely, the so-called asymmetric MOM scheme~\cite{Athenodorou:2016oyh,Boucaud:2017obn,Aguilar:2020yni,Aguilar:2021lke,Aguilar:2021okw,Ferreira:2023fva}. The latter is defined by the prescriptions
\be 
\Delta^{-1}(\mu^2) = \mu^2\,, \qquad F(\mu^2) = 1 \,, \qquad \Ls(\mu^2) = 1 \,,
\ee
where we choose $\mu = 4.3$~GeV as renormalization point. The corresponding value for the coupling is $g^2 = 4\pi\alpha_s$, with $\alpha_s(4.3 \text{ GeV}) = 0.27$, as determined in the lattice study of \cite{Boucaud:2017obn}. Within this renormalization scheme, the same SDE analysis of Refs.~\cite{Aguilar:2022thg,Ferreira:2023fva} used to determine $B_1(r,p,q)$ also yields the value ${\widetilde Z}_1 = 0.9333$.

\item Finally, $\IW(q,r,p)$ is a particular transverse projection of the three-gluon vertex, namely
\begin{align}
  \IW(q,r,p) &:= \frac 1 2 (q-r)^\nu {\overline \g}^{\alpha}_{\alpha\mu}(q,r,p) \,,
\label{eq:IWdef}
\end{align} 
which encodes the total contribution of ${\overline \g}^{\alpha}_{\alpha\mu}(q,r,p)$ to the SDE governing $\w(r^2)$. Note that the Bose symmetry of ${\overline \g}^{\alpha}_{\alpha\mu}(q,r,p)$ implies
\be 
\IW(q,r,p) = \IW(r,q,p) \,.
\ee 

\end{enumerate}

In a series of previous works~\cite{Aguilar:2020yni,Aguilar:2020uqw,Aguilar:2021uwa}, the ingredient appearing in \1eq{W_conts} that represented the largest uncertainty was $\IW(q,r,p)$. Since lattice results for the general kinematics three-gluon vertex were not available then, in those references $\w(r)$ had been approximated by various \emph{Ans\"atze} based on the Ball-Chiu construction of the three-gluon vertex~\cite{Ball:1980ax,Aguilar:2019jsj}. Recently, general kinematics lattice data for ${\overline \g}_{\alpha\mu\nu}(q,r,p)$ became available~\cite{Pinto-Gomez:2022brg,Pinto-Gomez:2022qjv,Pinto-Gomez:2023lbz}, prompting a more accurate determination of $\w(r)$.

Remarkably the general kinematics lattice results of Refs.~\cite{Pinto-Gomez:2022brg,Pinto-Gomez:2022qjv,Pinto-Gomez:2023lbz,Aguilar:2022thg}, as well as some continuum studies~\cite{Eichmann:2014xya,Blum:2014gna,Huber:2016tvc}, revealed that a compact expression provides a rather accurate approximation for the transversely projected three-gluon vertex. Specifically,
\be 
{\overline \g}_{\alpha\mu\nu}(q,r,p) \approx {\overline \g}^0_{\alpha\mu\nu}(q,r,p) \Ls(s) \,, \qquad s^2 := ( q^2 + r^2 + p^2 )/2 \,, \label{eq:compact}
\ee
where ${\overline \g}^0_{\alpha\mu\nu}(q,r,p)$ denotes the tree-level form of ${\overline \g}_{\alpha\mu\nu}(q,r,p)$. 

In \1eq{eq:compact} the sole dynamical ingredient is the soft gluon form factor, $\Ls(r)$, of \fig{fig:Lsg}, which now appears evaluated at the Bose-symmetric combination of momenta given by $s$. Note that general kinematics form factors of the three-gluon vertex are expected to depend on three Lorentz scalars. The fact that in \1eq{eq:compact} the form factor depends only on $s$, whose values define planes in the coordinate system $(q^2,r^2,p^2)$, has been termed \emph{planar degeneracy}~\cite{Pinto-Gomez:2022brg}.

Using the planar degeneracy approximation of \1eq{eq:compact} into \1eq{eq:IWdef}, we find a similarly compact expression for $\IW(q,r,p)$, namely
\be 
  \IW(q,r,p) \approx \IW^0(q,r,p)\Ls(s) \,,
\label{eq:IWcompact}
\ee
where $\IW^0(q,r,p)$ is the tree-level value of $\IW$, given by
\be 
\IW^0(q,r,p) := \frac{2f(q,r)}{p^2}\left[ 2 q^2 r^2 - (q^2 + r^2)(q\cdot r) - (q\cdot r)^2\right] \,. \label{IW0}
\ee 
The \1eq{eq:IWcompact} provides us with a baseline for computing $\w(r)$ accurately and expeditiously. 

In order to carry out the integrations over the whole momentum space in \1eq{W_diags}, we employ fits for the lattice data for $\Delta(r)$ and $F(r)$ from~\cite{Aguilar:2021okw} and for the $\Ls(r)$ of \cite{Aguilar:2019uob} which are constructed to reproduce the one-loop anomalous dimensions of these functions for large momenta. These fits are given in Appendix C of~\cite{Aguilar:2021uwa} and are all renormalized consistently in the asymmetric MOM scheme~\cite{Aguilar:2020yni,Ferreira:2023fva}. 

\begin{figure}[H]
\centering
\includegraphics[width=0.475\columnwidth]{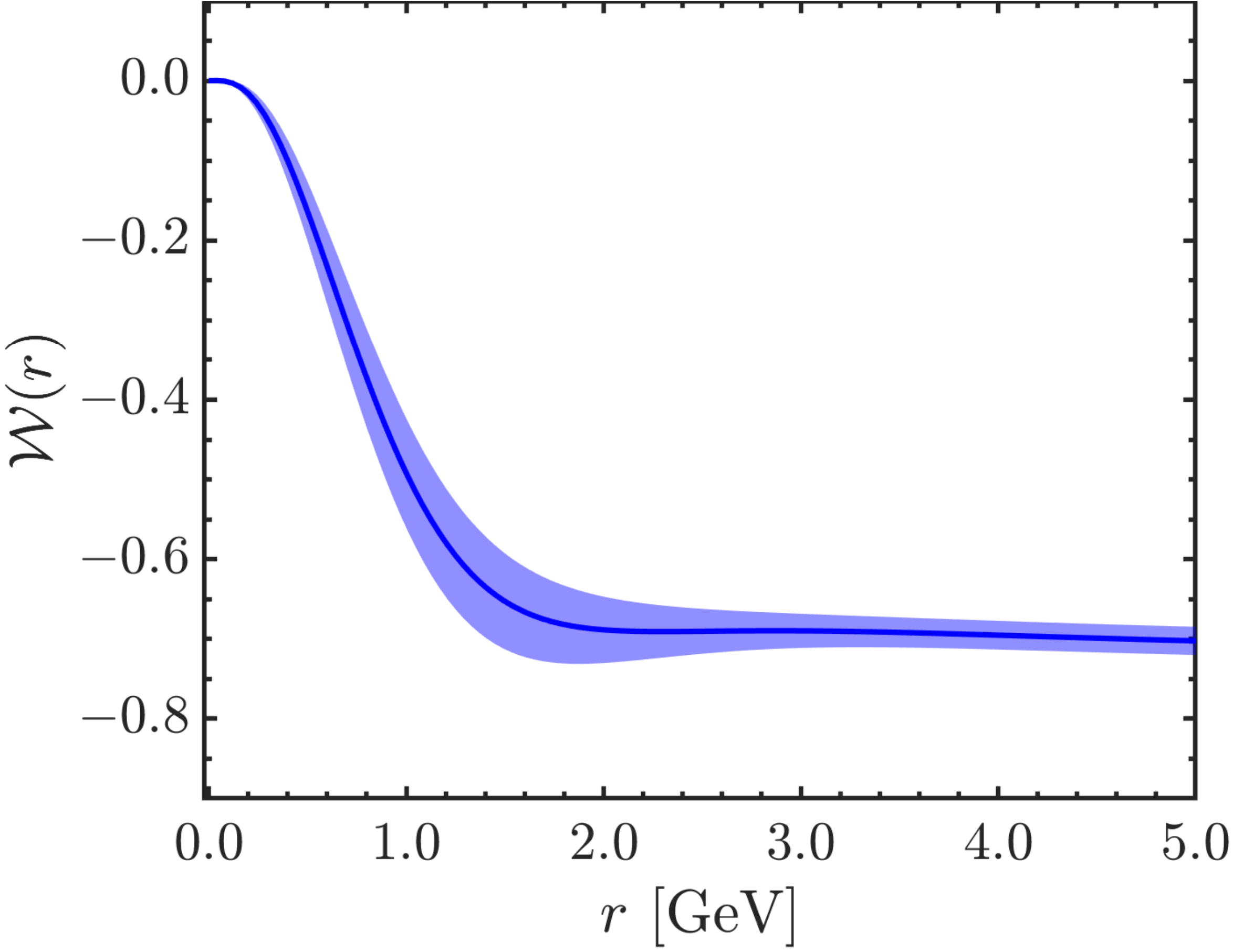} 
\includegraphics[width=0.475\columnwidth]{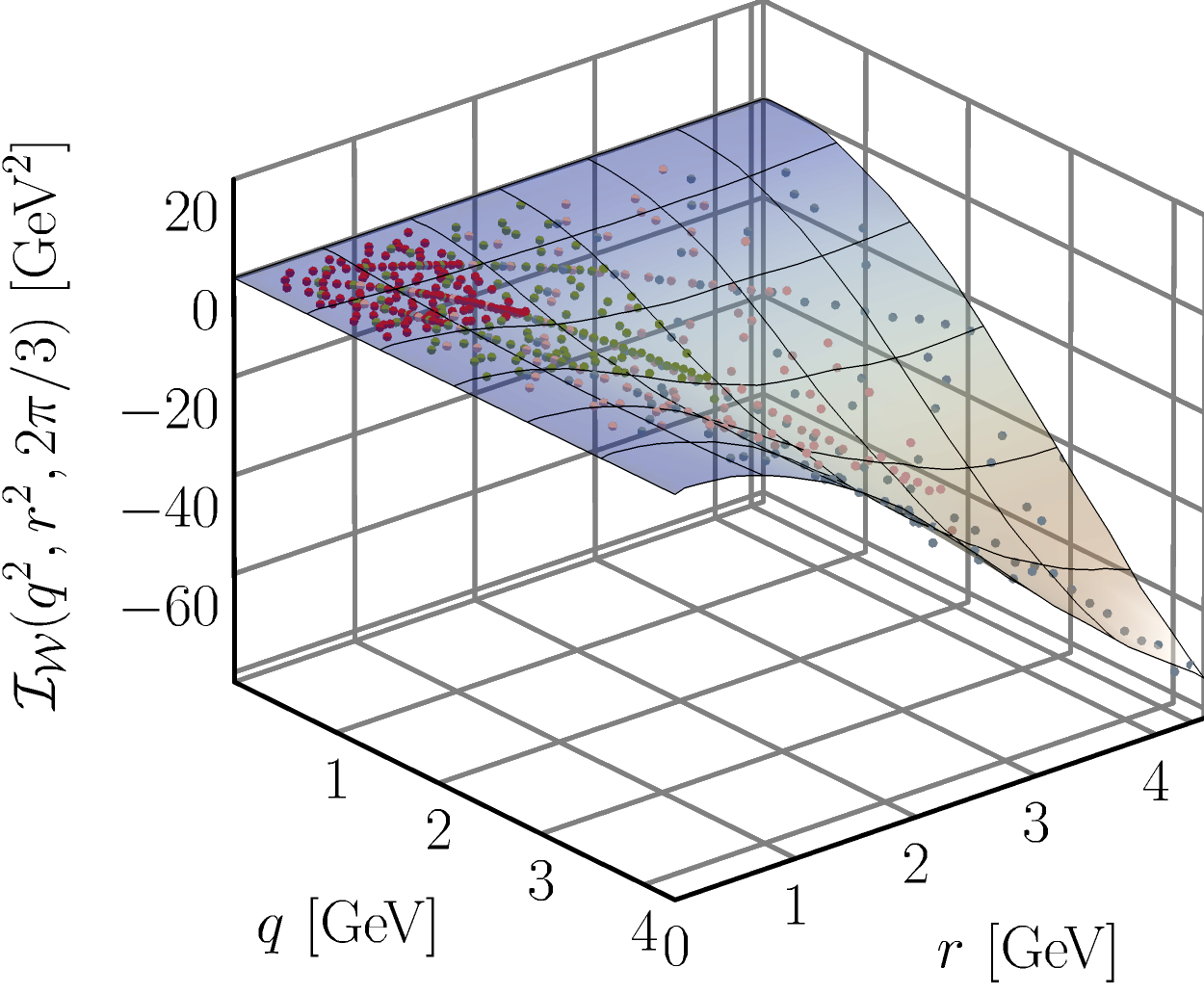} 
\caption{ Left: $\w(r)$ obtained using the planar degeneracy approximation of \1eq{eq:IWcompact} for the three-gluon vertex (blue solid curve) together with uncertainty estimate (blue band) obtained by using the Neural Network predictor for $\IW$. Right: Lattice results (points) for $\IW(q,r,p)$, when the momenta $q$ and $r$ are at an angle of $2\pi/3$. The surface shows the result of the Neural Network predictor trained on the data points. Another value for the angle can be seen in Fig. 4 of Ref.~\cite{Aguilar:2022thg}. }
\label{fig:W}
\end{figure}

Using the above ingredients, combined with the results for $B_1(r,p,q)$, ${\widetilde Z}_1$ and $\alpha_s$ mentioned in items $(i)$ and $(ii)$ above, we evaluate the Euclidean form of \1eq{W_diags} to obtain $\w(r)$. The result is shown as the blue solid curve in the left panel of \fig{fig:W}.

%%%%%%%%%%%%%%%%%%%%%%%%%%%%%%%%%%
% Figure 8 - W result
%%%%%%%%%%%%%%%%%%%%%%%%%%%%%%%%%%
%%%%%%%%%%%%%%%%%%%%%%%%%%%%%%%%%%

At this point, it is important to quantify the errors introduced in $\w(r)$ by the use of the approximate form of the three-gluon vertex given in \1eq{eq:compact}. To that end, the projection $\IW(q,r,p)$ has been computed \emph{directly} through lattice simulation in Ref.~\cite{Aguilar:2022thg}. Results for various lattice setups are shown as points in the right panel of \fig{fig:W}. These data correspond to momenta $q$ and $r$ at an angle of $2\pi/3$ and with arbitrary magnitudes. For different angles, the $\IW(q,r,p)$ is qualitatively similar.

In order to employ the lattice results for $\IW(q,r,p)$ into the SDE of $\w(r)$, we need a smooth interpolant. Since the data points depend on three kinematic variables (the magnitudes of two momenta and the angle between them), it is difficult to come up with a functional form that fits them accurately. Moreover, since the data is noisy, standard interpolants such as splines are unsuitable.

A reliable method to interpolate the general kinematics $\IW(q,r,p)$ consists of training a Neural Network predictor on the lattice data~\cite{Aguilar:2022thg}. To this end, we randomly selected one-third of the 335\,628 lattice points for $\IW(q,r,p)$ as a training set. The data was then fed into the Mathematica routine ``Predict'', with the option ``Neural Network'', which outputs a smooth predictor function. The remaining 223\,725 lattice data points were then used to confirm the accuracy of the resulting interpolant, by verifying that the predicted values were always within one standard deviation of the actual lattice results~\cite{Aguilar:2022thg}. 

In the right panel of \fig{fig:W}, the Neural Network predictor for $\IW(q,r,p)$ is represented by the color-mapped surface, which is compared to the full set of lattice data for the angle between $q$ and $r$ set at $2\pi/3$. In that figure, the accuracy and smoothness of the Neural Network result are clearly seen.

The Neural Network predictor for $\IW(q,r,p)$ can then be used directly into \1eq{W_diags}, as an alternative method for computing $\w(r)$. Quite remarkably, the results for $\w(r)$ computed with this method and those obtained from the planar degeneracy approximation of \1eq{eq:IWcompact} differ by only $2.5\%$~\cite{Aguilar:2022thg}. Combining this estimate of the systematic error with propagated statistical error of $\Ls(r)$~\cite{Aguilar:2022thg} we obtain a total error budget for $\w(r)$, which is represented as the blue band shown in the left panel of \fig{fig:W}.

%--------------------------------------------------
\section{Determination of the displacement amplitude from Lattice inputs}\label{Cdet}

Now we are in position to determine $\Cfat(r)$ from the WI displacement, \ie through \1eq{centeuc}.

\begin{figure}[H]
\centering
\includegraphics[width=0.475\columnwidth]{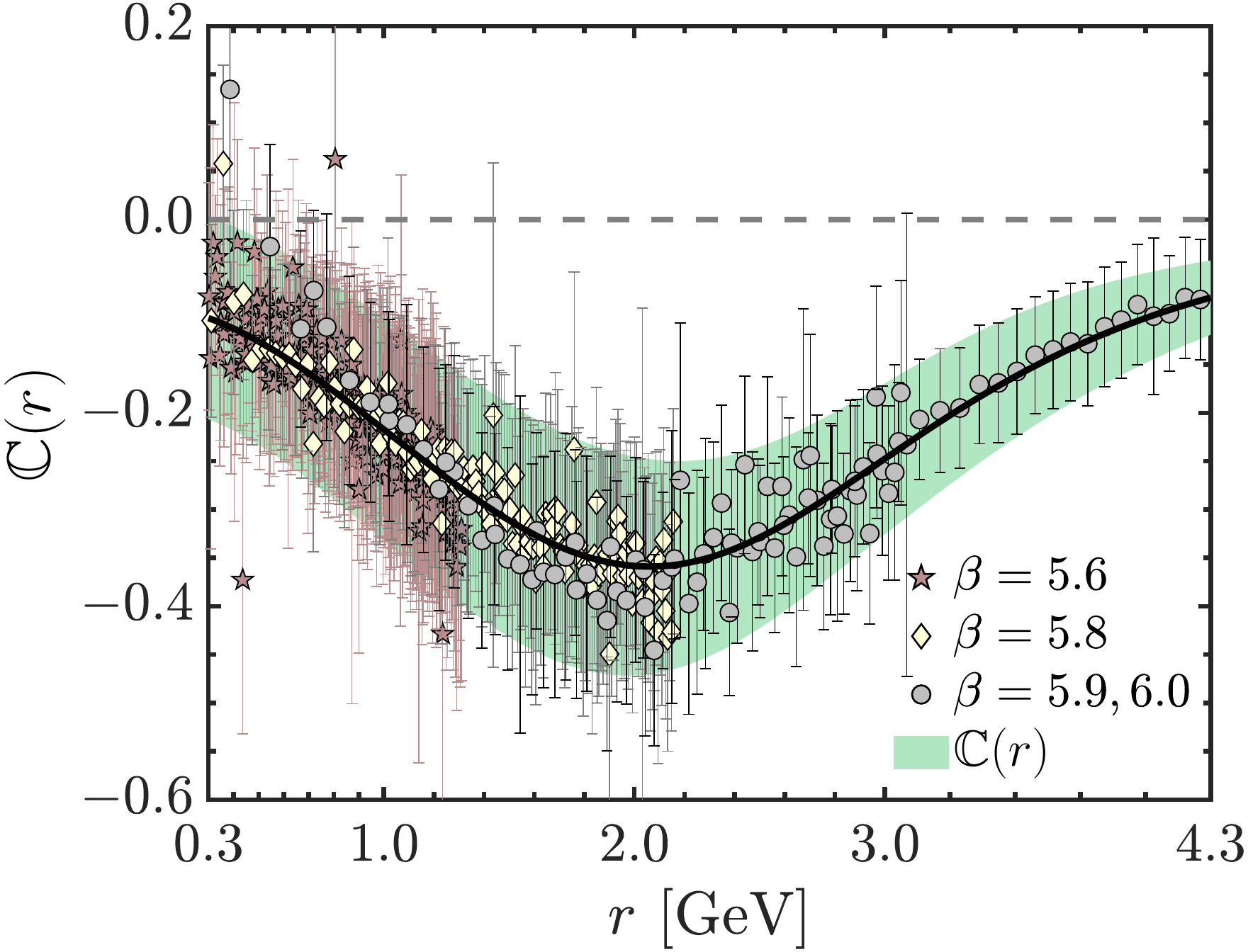}\hfil\includegraphics[width=0.475\columnwidth]{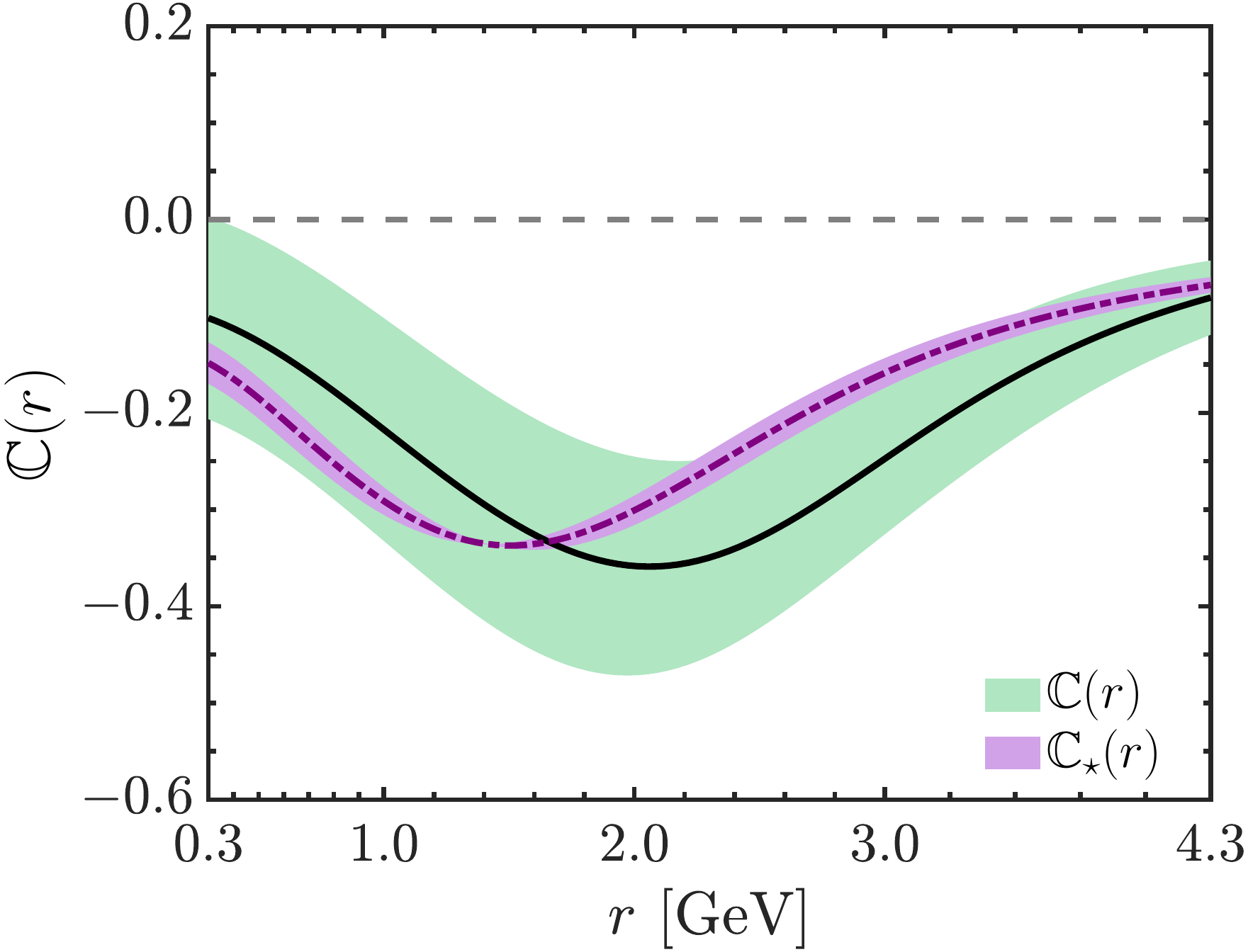}

\caption{ Left: Result for $\Cfat(r)$ obtained from the Ward identity (WI) displacement, \ie \1eq{centeuc}. The black continuous line results from using the lattice-driven SDE result for $\w(r)$, shown in the right panel of \fig{fig:W}, and fits for $\Delta(r)$, $\Ls(r)$ and $F(r)$. Using for $\Ls(r)$ the lattice data of Ref.~\cite{Aguilar:2021lke} directly yields the points. The green band emphasizes the typical size of the error estimate for $\Cfat(r)$ and is obtained by fitting the upper and lower bounds of the error bars of the points. Right: Comparison of the $\Cfat(r)$ (black line and green band) of the left panel to the BSE prediction, $\Cfat_\star(r)$, (purple dot-dashed and error band) of Ref.~\cite{Aguilar:2021uwa}.
}
\label{fig:Cfat}
\end{figure}

Combining the blue curve for $\w(r)$ of \fig{fig:W} with the aforementioned fits for $\Delta(r)$, $F(r)$ and $\Ls(r)$ into \1eq{centeuc}, we obtain for $\Cfat(r)$ the black solid curve in the left panel of \fig{fig:Cfat}. The points in the same panel show the result for $\Cfat(r)$ obtained by using \emph{directly} in \1eq{centeuc} the lattice data points of \cite{Aguilar:2021lke} for $\Ls(r)$, instead of a fit.

%%%%%%%%%%%%%%%%%%%%%%%%%%%%%%%%%%
% Figure 9 - C lattice result
%%%%%%%%%%%%%%%%%%%%%%%%%%%%%%%%%%
%%%%%%%%%%%%%%%%%%%%%%%%%%%%%%%%%%

The statistical significance of the above result for $\Cfat(r)$ can be quantified by comparing it to the null hypothesis, namely $\Cfat = \Cfat_0 = 0$. To this end, we compute the $\chi^2$ of our points for $\Cfat(r)$, with the null hypothesis taken as the estimator of the data, \ie
\be
\chi^2 = \sum_{i=1}^{n_r} \frac{\left[\Cfat(r_i)-\Cfat_0(r_i)\right]^2}{\epsilon_{\s{\Cfat(r_i)}}^2} = 2\,630\,. \label{chi2_def}
\ee 
In the above equation, $\epsilon_{\s{\Cfat(r_i)}}$ denotes the error estimate of $\Cfat(r_i)$ (the error bars in \fig{fig:Cfat}). The sum is performed over the \mbox{$n_r = 515$} indices $i$ such that $r_i \in [0.3, 4.3]$ GeV. 

From the result in \1eq{chi2_def}, we can compute the probability, $P_{\Cfat_0}$, that our result for $\Cfat(r)$ is consistent with the null hypothesis. Denoting by $\chi_{\rm \s{PDF}}^2(n_r,x)$ the $\chi^2$ probability distribution function with $n_r = 515$ degrees of freedom, we obtain~\cite{Aguilar:2022thg}
\be 
P_{\Cfat_0} = \int_{\chi^2=2\,630}^\infty \chi_{\rm \s{PDF}}^2(515,x) dx = \left.\frac{\Gamma(n_r/2,\chi^2/2)}{\Gamma(n_r/2)}\right\vert_{n_r = 515}^{\chi^2=2\,630} = 7.3\times10^{-280}\,. \label{P_C0}
\ee 
The vanishingly small probability obtained in \1eq{P_C0} is to be understood as meaning that, in the absence of additional uncertainties or correlations in the data, the null hypothesis $\Cfat_0$ is completely excluded. Moreover, we point out that even if the error of \emph{every data point} for $\Cfat(r)$ was $95\%$ larger we could still discard $\Cfat_0$ at the $5\sigma$ confidence level.

The result for $\Cfat(r)$ obtained in this way can then be compared to the BSE prediction, $\Cfat_\star(r)$, of Ref.~\cite{Aguilar:2021uwa}, shown in the right panel of \fig{fig:BSE}. To this end, we first need to determine the overall scale and sign of $\Cfat_\star(r)$, which are left undetermined by the homogeneous nature of the BSE. 

Denoting by $\Cfat_{\s {\rm BSE}}(r)$ a solution of the homogeneous BSE, we first define 
\be 
\Cfat_\star(r) = b\,\Cfat_{\s {\rm BSE}}(r)\,,
\ee
with $b$ a constant.
Then, we determine the multiplicative constant $b$ by minimizing the $\chi^2$ measure for the discrepancy between $\Cfat$ and $\Cfat_\star$ as
\be
\chi^2_\star = \sum_{i} \frac{\left[\Cfat(r_i)-\Cfat_\star(r_i)\right]^2}{\epsilon_{\s{\Cfat(r_i)}}^2} \,. \label{chi2_star_def}
\ee 
The result of this procedure is the $\Cfat_\star(r)$  shown previously in the right panel of \fig{fig:BSE}. 

Next, in the right panel of \fig{fig:Cfat} we compare $\Cfat(r)$ and $\Cfat_\star(r)$ directly, finding a rather good agreement. The main difference is in the position of the minimum, which is shifted from $r = 1.93\substack{+0.09 \\ -0.06}$~GeV for $\Cfat(r)$ to $r=1.5\pm 0.1$ for $\Cfat_\star(r)$.

Finally, in addition to determining the scale and sign of $\Cfat_\star(r)$, the $\chi^2_\star$ measure of \1eq{chi2_star_def} allows us to perform a statistical analysis of the compatibility between the BSE prediction and the lattice result. Specifically, after setting the scale of $\Cfat_\star$ we obtain $\chi^2_\star = 258.5$, which is smaller than the number of degrees of freedom. Indeed, this value of $\chi^2_\star$ translates to a near unit probability,
\be 
P_{\Cfat_\star} = \left.\frac{\Gamma(n_r/2,\chi^2_\star/2)}{\Gamma(n_r/2)}\right\vert_{n_r = 515}^{\chi^2_\star=258.5} = 1 - 2.0\times 10^{-23} \,,
\ee 
of the points $\Cfat(r)$ being compatible with $\Cfat_\star(r)$~\cite{Ferreira:2023fva}.

%--------------------------------------------------
\section{Conclusion}\label{conc}

The displacement of the WIs by the formation of massless vertex poles is a distinctive feature of the Schwinger mechanism for gluon mass generation, which allows its verification from lattice QCD results. In the present work, we have used this framework to demonstrate that the three-gluon vertex in Yang-Mills SU(3) has such a pole. Indeed, our analysis of the WI displacement using lattice data unequivocally excludes the null hypothesis of a vanishing BS amplitude, $\Cfat = 0$. Instead, our results reveal an excellent agreement between the $\Cfat(r)$ derived from the WI and the BSE prediction, providing outstanding evidence for the occurrence of the Schwinger mechanism in QCD.

It is important to emphasize that while the present analysis was carried out in the simpler setting of pure Yang-Mills SU(3), the same ideas hold in the presence of dynamical quarks. In particular, the WI displacement for the three-gluon vertex retains exactly the same form as in \1eq{centeuc} in the unquenched case, for which lattice data for the propagators and vertices also exist~\cite{Bowman:2004jm,Kamleh:2007ud,Ayala:2012pb,Cui:2019dwv,Aguilar:2019uob}. Indeed, a study is already underway to investigate the Schwinger mechanism poles in the presence of quarks and should be reported soon.

Finally, in the present work, we have focused entirely on the massless pole content of the three-gluon vertex. However, once the Schwinger mechanism is active, it is expected that massless poles appear in various vertices~\cite{Aguilar:2016vin}, since the different vertices are connected to one another through the SDEs. As such, other important signals of the Schwinger mechanism may be present in functions such as the ghost-gluon kernel, and the quark-gluon and four-gluon vertices, which are currently under investigation as well.

\backmatter

%--------------------------------------------------
\bmhead{Acknowledgments}

The author thanks A.C. Aguilar, J. Papavassiliou, C.D. Roberts, and J. Rodr\'iguez-Quintero for the collaborations.

%--------------------------------------------------
\section*{Declarations}

\bmhead{Funding}

M.N.F. is supported by the grant PID2020-113334GB-I00 and the contract CIAPOS/2021/74, from the Spanish MICINN and the Generalitat Valenciana, respectively.

\bmhead{Competing interests}

The author has no relevant financial or non-financial interests to disclose.


\end{document}